\documentclass[12pt]{article}
\usepackage{latexsym,epsfig,graphicx,amsmath,amssymb,amscd,undertilde,multirow,chicago,psfrag,paralist,dsfont}
\usepackage[titletoc]{appendix}
\newcommand{\thetavec}{{\boldsymbol{\theta}}}

\newcommand{\alphavec}{{\boldsymbol{\alpha}}}
\newcommand{\omegavec}{{\boldsymbol{\omega}}}

\newcommand{\betavec}{{\boldsymbol{\beta}}}

\newcommand{\N}{{\textrm{N}}}

\newcommand{\nor}{\text{nor}}

\begin{document}

\title{Big Data and Reliability Applications: The Complexity Dimension}
\author{Yili Hong and Man Zhang\\
Department of Statistics\\
Virginia Tech\\
Blacksburg, VA 24061\\
\and
William Q. Meeker\\
Department of Statistics\\
Iowa State University\\
Ames, IA 50011\\
}

\date{\today}

\maketitle
\begin{abstract}
Big data features not only large volumes of data but also data with complicated structures. Complexity imposes unique challenges in big data analytics. Meeker and Hong (2014, Quality Engineering, pp. 102-116) provided an extensive discussion of the opportunities and challenges in big data and reliability, and described engineering systems that can generate big data that can be used in reliability analysis. Meeker and Hong (2014) focused on large scale system operating and environment data (i.e., high-frequency multivariate time series data), and provided examples on how to link such data as covariates to traditional reliability responses such as time to failure, time to recurrence of events, and degradation measurements. This paper intends to extend that discussion by focusing on how to use data with complicated structures to do reliability analysis. Such data types include high-dimensional sensor data, functional curve data, and image streams.  We first provide a review of recent development in those directions, and then we provide a discussion on how analytical methods can be developed to tackle the challenging aspects that arise from the complexity feature of big data in reliability applications. The use of modern statistical methods such as variable selection, functional data analysis, scalar-on-image regression, spatio-temporal data models, and machine learning techniques will also be discussed.

\textbf{Key Words:} Clustering; Degradation Data; Functional Data; Machine Learning; Reliability Prediction; Spatio-temporal Data.

\end{abstract}

\newpage

\section{Introduction}
\subsection{Background}
Technological advancements have fundamentally changed the way that data are collected, causing the arrival of the big data era. Big data features not only large data volumes and high speed in data collection, but also data with complicated structures. In many applications, complexity imposes unique challenges in big data analytics. As pointed out by \citeN{DavidMSteinberg2016}, technological development and industrial advancement continue to generate challenging problems that will require new statistical methods, which is also true for research on the area of reliability analysis. \citeN{MeekerHong2014} provided an extensive discussion of the opportunities and challenges in big data and reliability. A large number of engineering systems that can generate big data were discussed. Those engineering systems can generate large-scale system operating and environment data (i.e., high-frequency multivariate time series data), which can be used in reliability analysis. \citeN{MeekerHong2014} also provided examples on how to link the large-scale system operating and environmental data to traditional reliability responses such as time to failure, time to recurrence of events, and degradation measurements.

The main objective of this paper is to extend the discussion in \citeN{MeekerHong2014} by focusing on the use of data with complicated structures to do reliability analysis. Traditional data types that have been used for reliability data analysis include lifetime data and degradation data. With the advancement of technology, various data types other than lifetime data and degradation can be collected from engineering systems and reliability tests. Those data types include high-dimensional sensor data, functional curve data, and image streams. With appropriate analytical methods, it is possible that those new data types can be used to provide reliability information for the systems and products. Instead of using traditional tools from survival analysis for lifetime data and nonlinear mixed effects models for degradation data, several modern statistical methods can be tailored and integrated to analyze reliability data with complicated structures. Those potentially useful modern statistical methods include variable selection, functional data analysis, scalar-on-image regression, spatio-temporal data modeling, and machine learning techniques.

In this paper, we first provide a review of recent developments in modeling and analysis of reliability data with complicated structures. We then provide a discussion on how analytical methods can be developed to tackle the challenging aspects that arise from the complexity features of big data in reliability applications.

\subsection{Data Complexity and Reliability Applications}
The common characteristics of big data are described by the three V's (e.g., \citeNP{Zhang2015}), which represent volume, velocity, and variety.  ``Volume'' refers to the large amount of data that can be collected, when there is continuous tracking of the past. ``Velocity'' means that data are recorded with high frequency and are available in real time. ``Variety'' refers to that data arrive in various formats such as text, images, audio, and video. The system operating and environmental (SOE) data presented in \citeN{MeekerHong2014} represents the volume and velocity of the big data features in reliability, because the SOE data track how the product being used and under which environments being used in real time.  The ``variety'' feature is related to data complexity. Data complexity can include high dimensionality, complex relationships, and many other complications in a dataset.

While complexity in big data presents tremendous challenges in modeling and analysis, it also provides opportunities to develop new statistical methods. This is also true in reliability applications. The types of data that are used for reliability analysis evolve over time. Lifetime data have had a long history being used to provide reliability information. In the past decades, degradation data have also been used to provide reliability information. With the arrival of big data technology, we see new data types become available for reliability analysis. Even though some of these data types have existed in the past, the current big data wave makes the cost of collection low and provides motivation to analyze those data with complex structures. Examples of such data include:

\begin{inparaitem}

\item \emph{Multivariate time series data}:  multi-channel sensor data that are collected at regular intervals is an example of this type of data. The SOE data presented in \citeN{MeekerHong2014} provide detailed examples.

\item \emph{Functional data}: data collected by equipment such as a spectrophotometer display a functional curve. High-frequency sensor data can also be treated as functional data. Functional data can also be available over time, which we call longitudinal functional data.

\item \emph{Image data and streams}: image data can be collected by infrared cameras, or other more advanced equipment such as a scanning electron microscope (SEM) to characterize system status or material properties. When images are taken over time, image streams are obtained.

\item \emph{Other types of Unstructured data}: Data types such as text data and audio data can also provide reliability information in some applications. For example, text data in a warranty database can provide information about product maintenance. Audio data can be used for the monitoring of system operating status.

\end{inparaitem}

\subsection{Overview}
We intend to discuss the opportunities that big data bring in the following areas of reliability analysis, especially from data a complexity point of view.

\begin{inparaitem}
\item  \emph{New trends in degradation data analysis}: Section~\ref{sec:degradation.analysis} reviews recent developments in degradation data analysis and discusses some new directions for methodological research in degradation data analysis.

\item \emph{Tackling new types of covariates}: Section~\ref{sec:tackling.covariates} discusses some opportunities in tackling new types of covariates in reliability analysis. We focus on how to automatically select important covariates for reliability prediction and how to incorporate functional and image type of predictors for reliability metrics.

\item \emph{Applications of machine learning techniques}: Section~\ref{sec:machine.learning.rel.pred} discusses applications of several machine learning techniques in reliability analysis. These include clustering of events, deep learning for predictions, and text analytics for predictions.

\item \emph{Emerging application areas of reliability}:  Section~\ref{sec:new.areas} discusses several emerging application areas where reliability analysis techniques can be used.

\end{inparaitem}
Finally, Section~\ref{sec:conclusions} contains some concluding remarks.

\section{Degradation Analysis}\label{sec:degradation.analysis}
Degradation data have been widely used to conduct reliability prediction and system health assessment. \citeN{LuMeeker1993} novelly used degradation measurements to assess reliability. Such data are typically repeated measurements of a degradation index (e.g., the depth of tire tread) over time. Statistical methods to analyze degradation data are mainly based on two types of models: the general paths models (e.g., \citeNP{Nelson1990}, Chapter~13 of \citeNP{meekerescobar1998}, and \shortciteNP{EscobarMeekerKuglerKramer2003}), and stochastic models (e.g., \citeNP{Whitmore1995}, and \citeNP{ParkPadgett2005}).

Recent developments of stochastic models for repeated measurement degradation includes \citeN{WangXu2010}, \citeN{YeChen2014}, and \citeN{Peng2016} for inverse Gaussian process models. \citeN{ZhouSerbanGebraeel2011} and \shortciteN{Zhouetal2014} used functional data analysis techniques to model the degradation paths. For destructive degradation test data, \shortciteN{Xieetal2016} developed a semiparametric model to describe the degradation path. For products and systems used in the field, the units are typically subject to time-varying usage and environmental conditions. With the advancement of data collection technology, one can record such time-varying covariates, which is referred to as degradation data with dynamic covariates. Recent development on statistical methods for degradation data with dynamic covariates can be found at \shortciteN{HongDuanetal2015} and \shortciteN{XuHongJin2015}. \citeN{ChenYe2018} considered uncertainty quantification for degradation models. However, many other research opportunities arise due to the arrival of big data, especially related to the complexity features.

\subsection{Degradation Index Construction}
\begin{figure}
\begin{center}
\includegraphics[width=0.65\textwidth]{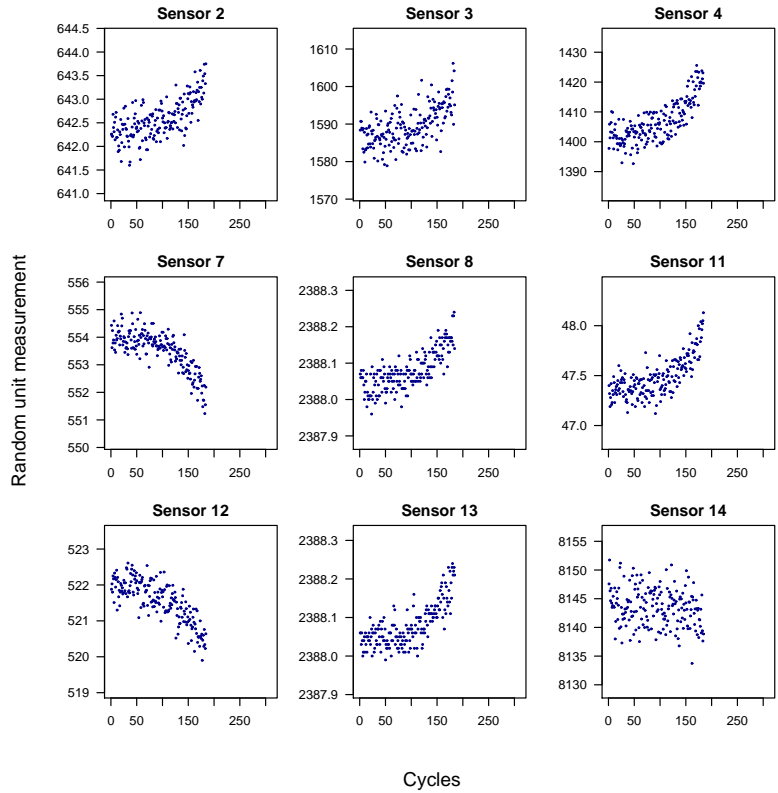}
\caption{An example of multi-channel sensor data from the NASA jet engine simulation data. The plot shows the signal measurement as a function of cycles from a subset of variables from one randomly selected unit.}\label{fig:jet.dat}
\end{center}
\end{figure}

Most of the existing research on degradation data modeling assumes that the degradation index for a product or system is well defined. Modern sensor technology allows one to collect multi-channel sensor data that are related to the underlying degradation process. However, any single channel may not be sufficient to represent the underlying degradation process. Figure~\ref{fig:jet.dat} shows an example of multi-channel sensor data from the NASA jet engine simulation data (e.g., \citeNP{SaxenaGoebel2008}). Without a well-defined index, most of existing methods will not be applicable. Thus, constructing an appropriate degradation index is a fundamental step in degradation modeling. \citeN{LiuGebraeelShi2013} proposed a data-level fusion model for developing composite health indices for degradation modeling and prognostic analysis. \citeN{FangPaynabarGebraeel2017} studied a multi-stream sensor fusion-based prognostics model for systems with a single failure mode. \shortciteN{Chehadeetal2018} considered a data-level fusion approach for degradation index building under multiple failure modes.

Here we briefly discuss a general approach for degradation index building based on an additive-nonlinear model with variable selection. The approach is more flexible than a linear combination of sensor signals, and the approach can automatically select the most informative variables to be used in the degradation index. Let $x_i(t)=[x_{i1}(t), \cdots, x_{ip}(t)]'$ be the multivariate measurements for unit $i$ at time $t$, $i=1,\cdots, n$. Here, $p$ is the number of sensor channels and $n$ is the number of units. The degradation index is built as
$$z_i(t)=\sum_{j=1}^{p}f_j[x_{ij}(t); \betavec_j],$$
where $f_j$ is the contribution function for variable $j$. The functional forms of the $f_j$ are expressed as a linear combination of spline bases with parameter vector $\betavec_j$. Methods for construction of the spline bases for $f_j$ can be found in \citeN{Meyer2008}. The procedure to estimate the $f_j$ functions is briefly discussed below.

Let $t_{ik}$, $k=1, \cdots, n_i$, be the time point where the measurements are taken. Let $t_i=t_{in_i}$ be the last measurement time point and let $\delta_i$ be an event indicator. The indicated events may not be failures but can be events where the product performance requirements cannot be met for specific types of products. The events typically indicate that the degradation has progressed above a certain threshold that is unknown. For those units with events at time $t_i$ (i.e., $\delta_i=1$), the degradation index $z_i(t_i)$ should be near to the threshold with small variation. That is, the goal is to minimize the variation of $z_i(t_i)$ for those events. In particular, the objective is to minimize $$\min_{\betavec_j,\,j=1,\cdots,p}\sum_{\{i:\,\,\delta_i=1\}}[z_i(t_i)-\bar{z}]^2,$$
where $\bar{z}=\sum_{\{i:\,\,\delta_i=1\}}{z_i(t_i)}/\sum_{i=1}^{n}{\delta_i}$ is the mean degradation level.

A group lasso penalty function $\lambda_{1}\sum_{j=1}^{p}||\betavec_j||$ (e.g., \citeNP{YuanLin2006}) is used to select the most informative variables into the degradation index. Here $||\cdot||$ is the $L_{2}$ norm and $\lambda_1$ is a tuning parameter. Note that the degradation index should be monotonically increasing. To impose this constraint, a penalty for non-monotonicity is introduced. In particular, the penalty term is
$$\lambda_{2}\sum_{i=1}^{n}\sum_{k=1}^{n_i}\left[z_i\left(t_{i,k-1}\right)-z_i\left(t_{i,k}\right)\right]_{+},$$
where $[x]_{+}=x+c$, if $x>0$, and $[x]_{+}=0$, if $x<0$, and $\lambda_2$ is a tuning parameter. Here, the constant $c$ is used to impose strict monotonicity. In summary, the overall problem is formulated as follows:
\begin{align}\label{eqn:deg.index.fun}
\min_{\betavec_j,\,j=1,\cdots,p} \sum_{\{i: \delta_i=1\}}{[z_i(t_i)-\bar{z}]}^2 +\lambda_{1}\sum_{j=1}^{p}{||\betavec_j||}+\lambda_2\sum_{i=1}^{n}\sum_{k=1}^{n_i}\left[z_i\left(t_{i,k-1}\right)-z_i\left(t_{i,k}\right)\right]_{+}.
\end{align}
The coordinate decent algorithm in \citeN{TibshiraniTaylor2011} can be used to solve \eqref{eqn:deg.index.fun}.

\subsection{Multivariate Degradation Modeling}
\begin{figure}
\begin{center}
\begin{tabular}{cc}
\includegraphics[width=0.45\textwidth]{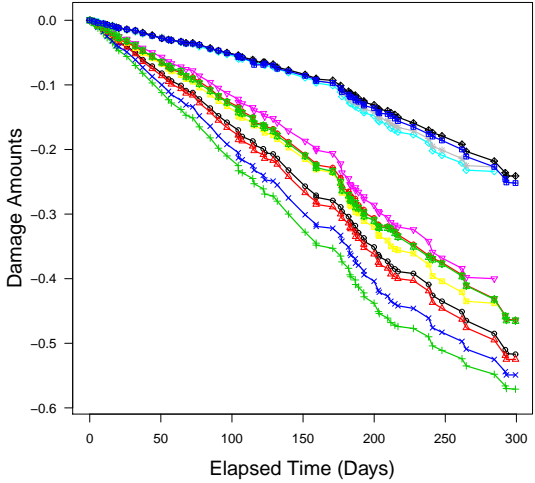}&
\includegraphics[width=0.45\textwidth]{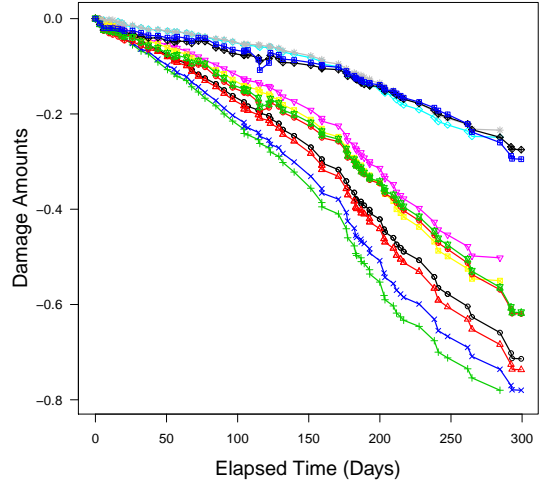}\\
(a) Damage 1250 cm$^{-1}$ & (b) Damage 1510 cm$^{-1}$\\
\end{tabular}
\caption{Examples of degradation data with two characteristics from the NIST coating data. Each connected line shows the observed degradation path for a unit and the dots show points in time the measurements were taken.}\label{fig:nist.coating.dat}
\end{center}
\end{figure}

Most of the existing framework for degradation data modeling focuses on a single degradation characteristic. In some applications, however, systems may have multiple characteristics that degrade simultaneously, and one can track those degradation characteristics together. Figure~\ref{fig:nist.coating.dat} shows examples of degradation data with two degradation characteristics from the National Institute of Standards and Technology (NIST) coating data (\shortciteNP{Guetal2008}).

The literature for multivariate degradation modeling is sparse, mostly due to the lack of flexible models for such data. The multivariate Wiener process is one exception because the multivariate extension of the well-known Wiener process retains the independent increment and infinite divisibility properties (e.g., \citeNP{WhitmoreCrowderLawless1998}). Some papers have used a copula (e.g., \citeNP{Nelsen2006}) to model the joint distribution of the increments (e.g., \citeNP{PanBalakrishnan2011}, and \shortciteNP{Panetal2013}). However, direct modeling of the increments by using a copula model does not preserve the infinite divisibility property. Thus, it can be problematic (not well-defined) when one adds together the increments of the two consecutive intervals, because the distribution of the sum of the increments is no longer from the same class of models. \shortciteN{Sunetal2016a} and \shortciteN{Sunetal2016} considered multiple degradation characteristics under an accelerated degradation test setting. \shortciteN{Sietal2018} conducted reliability analysis for dynamic local deformation of materials under a multivariate degradation model.

Here, we discuss a general model structure based on a copula random effects model. In general, the stochastic model for the multivariate degradation measurement is
$$Y_{j} (t)=D_{j}(t)+\epsilon_{j}(t), \quad j=1,\cdots, p,$$
where $D_{j}(t)$ is the underlying stochastic process for the true degradation process of characteristic $j$, $\epsilon_{j}(t)$ is the error term, and $p$ is the number of degradation characteristics. The commonly-used Wiener process, gamma process, and inverse Gaussian process can be used to describe the stochastic behavior of $D_{j}(t)$. However, a new structure needs to be introduced into $D_{j}(t)$ in such way that it can capture a flexible dependence structure among the $p$ characteristics, while each marginal process is still a well-defined process. The copula random effects model meets the requirements.

In this paper, we use the Wiener process as an illustration. The copula random effects model can also be used for other processes such as the gamma process and the inverse Gaussian process. In particular, the new model structure is
\begin{align}\label{eqn:deg.copula.model}
&D_{j}(t)=D_{j}(t; x,\omega _{j})=\omega_{j}\Lambda_{j}(t; x)+\sigma_{j}B[\Lambda_{j}(t;x)],\quad j=1,\cdots, p,\\\nonumber
&\omegavec=(\omega_{1},\cdots ,\omega_{p} )'\sim C[F_{1}(\omega_{1} ),\cdots, F_{p}(\omega_{p})].
\end{align}
Here, $\Lambda_{j}(\,\cdot\,; x)$ is the shape function where $x$ represents covariate information, and $B(\cdot)$ is the standard Wiener process. The random vector $\omegavec$ is used to introduce the dependence structure among all stochastic processes.  Conditional on $\omegavec$, each $D_{j}(t)$ is a well-defined Wiener process. The function $C(u_1, \cdots, u_p)$ here is a copula function and the $F_{j}(\omega_j)$ functions are marginal cumulative distribution functions (cdf) for $\omega_{j}$, $j=1,\cdots p$. The commonly-used copulas, such as the Gaussian and Archimedean copulas can be used (e.g., \citeNP{Nelsen2006}). The commonly-used marginal distributions such as lognormal, the Weibull, and gamma can also be used. With the combination of the copula functions and the marginal distributions, a great deal of flexibility for the dependence structure can be achieved. Note that the model in \eqref{eqn:deg.copula.model} is also capable of incorporating covariate information $x$ into the model, for example, using a regression type of model. The model estimation can be done by using the Monte Carlo EM algorithm (e.g., \shortciteNP{Bedairetal2016}).

\subsection{Longitudinal Functional Data for Degradation Modeling}
\begin{figure}
\begin{center}
\includegraphics[width=0.9\textwidth]{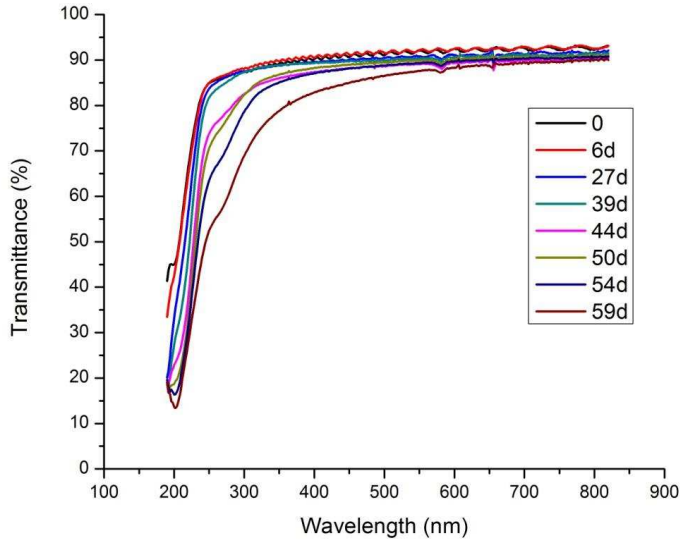}
\caption{Temporal changes of light transmittance spectra for an EVA sample under UV irradiation. The legend shows the number of days from the origin when the measurement was taken.}\label{fig:nist.optical.property.dat}
\end{center}
\end{figure}

As measurement instruments advance and data storage capacity increases, new types of data are becoming available for degradation analysis. Instead of generating one single measurement for characterizing a material property, some instruments will generate a functional curve to represent the property of a material. For example, transmission spectroscopy can measure the light transmittance as a function of wavelength. The change in the transmittance spectra will indicate the deterioration of a material property. Thus, the longitudinal measurements of the transmittance spectra can show the trend of the material property degradation over time.  Figure~\ref{fig:nist.optical.property.dat} shows the temporal changes of light transmittance spectra for an ethylene-vinyl acetate (EVA) sample under ultraviolet (UV) irradiation.

There is little work in reliability data analysis literature dealing with longitudinal functional data. In the functional data analysis literature, \citeN{ParkStaicu2015} developed longitudinal functional data analysis techniques to handle such data. Here we briefly describe the model in \citeN{ParkStaicu2015}. Let $Y_{ij}(s), s\in \Omega$ be the function curve for unit $i$ at time $t_{ij}$. Here $\Omega$ is the domain of the functional curve, and $t_{ij}$ is the time where the measurement are taken. In particular, the measurement $Y_{ij}(s)$ is modeled as
\begin{align}\label{eqn:deg.curve.fpca}
Y_{ij}(s)=\mu(s, t_{ij})+X_i(s, t_{ij})+\epsilon_{ij}(s).
\end{align}
The model in \eqref{eqn:deg.curve.fpca} has three major components. The term $\mu(s, t_{ij})$ is the mean structure at time $t_{ij}$, and $\epsilon_{ij}(s)$ is an error term. The individual to individual differences are described by the term,
\begin{align*}
X_i(s, t_{ij})=\sum_{k\geq 1} \xi_{ik}(t_{ij})\psi_k(s),
\end{align*}
which is a linear combination of a common basis function $\psi_k(s)$ with coefficients $\xi_{ik}(t_{ij})$. Note that $\xi_{ik}(t)$ is a function of $t$. Thus it is time dependent. The estimation method is mainly based on functional eigenvalue decompositions. However, for degradation data, one often assumes a monotonicity behavior among observed curves as the time gets larger. Also, one of the major goals of reliability analysis is to make predictions. Thus, monotonicity and extrapolation are the two major challenges that need to be addressed in order to make the model in \eqref{eqn:deg.curve.fpca} applicable in reliability applications.

\subsection{Spatio-temporal Data for Degradation Modeling}
\begin{figure}
\begin{center}
\includegraphics[width=0.8\textwidth]{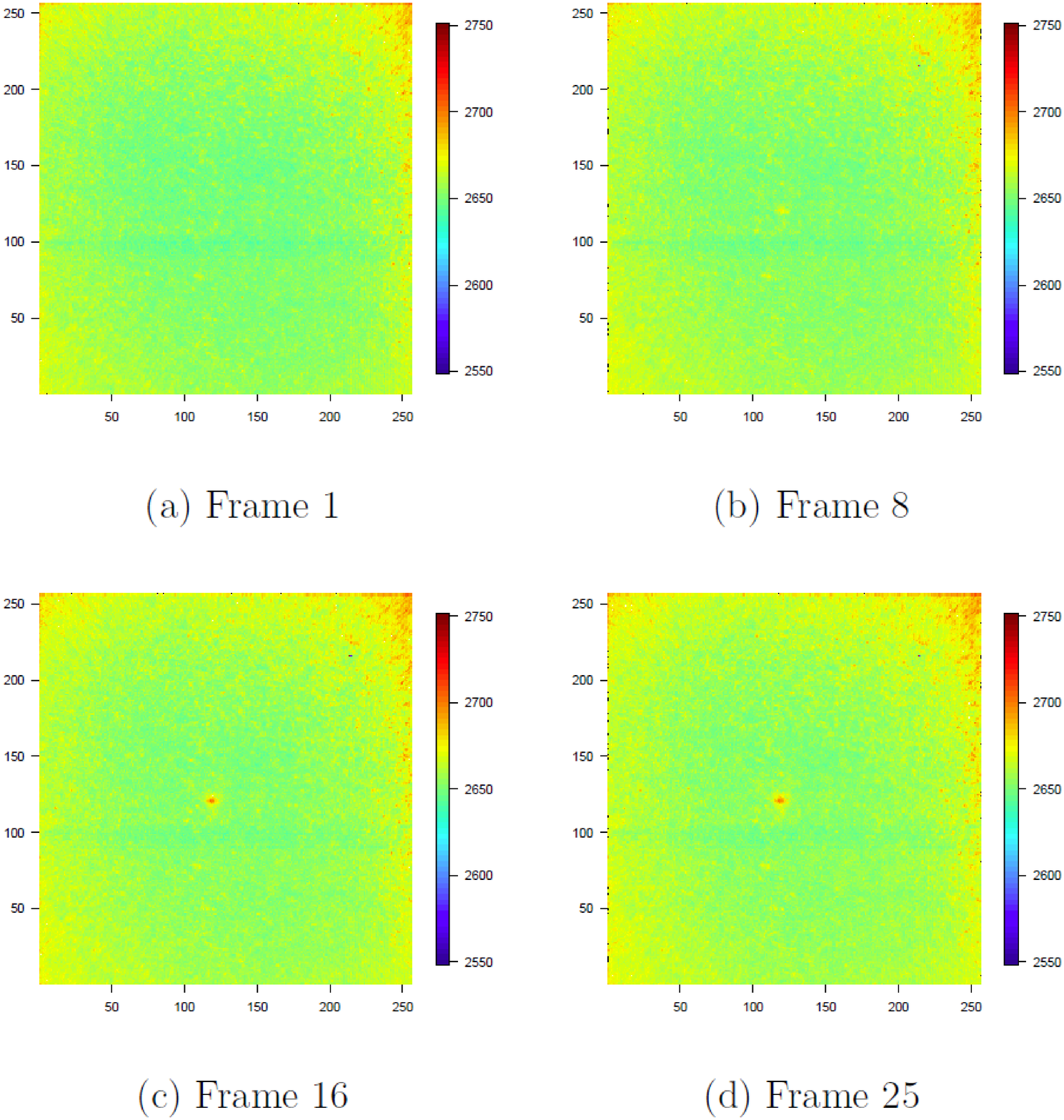}
\caption{Illustration of the evolution of a spatio-temporal process using the vibrothermography data. The color (darkness) shows the intensity. }\label{fig:sonic.movie.dat}
\end{center}
\end{figure}

In some applications, the measurements are taken over a spatial region for a period of time, which is related to the degradation process. We call this type of data as spatio-temporal degradation data. For example, infrared cameras can be used to track the change of a thermal field. Although it is not directly related to degradation, vibrothermography data from \citeN{GaoMeeker2012} in Figure~\ref{fig:sonic.movie.dat} illustrate the evolution of a process over a two-dimensional spatial region. From frame 8, one starts to see a signal in the middle of the region and the signal becomes evident in frame 16. Another example of spatio-temporal degradation is available in \citeN{LiuYeoKalagnanam2016}, who proposed to use spatio-temporal models to describe such types of data. They addressed challenges relating to spatial heterogeneity, the spatial propagation of degradation to neighboring areas, the anisotropic and space-time nonseparable covariance structure in a complex spatio-temporal degradation process. Other examples where spatio-temporal degradation data were generated include the spatial variation of a thermal field and microscopic characterization for the wear of tools. In silicon ingot manufacturing, equipment degradation will affect the spatial variation of the thermal field, and a change in the thermal field will affect the quality of the ingot. In a broaching process, the tool condition of the broach has high impacts on the quality of the final part. Images from a microscope are used to characterize the wear of the broaching tool over time.

Here we briefly describe a Bayesian framework that can be used for modeling and analysis of spatio-temporal degradation data. Assume that we have measurements over a spatial region $\Omega$ for a period $[0,\tau]$. We denote the spatio-temporal data by $D(s,t),\; s\in\Omega,\; t\in [0,\tau]$. Let $s_1,\cdots, s_m$ be the locations in the spatial domain where the measurements are taken, and let $m$ denote the number of locations. Let
$$D_{t}=[D(s_{1}, t), D(s_{2}, t),\cdots, D(s_{m},t)]',$$
where $s_{1}, s_{2}, \cdots, s_{m} \in \Omega,$ and $t\in [0,\tau ]$. Note that $D_{t}$ is a vector that contains all the observed measurements taken at the $m$ spatial locations at time $t\in [0,\tau]$. One can model $D_{t}$ via a Bayesian hierarchical model (e.g., \citeNP{CressieWikle2011}). We use the following model for the measurements,
\begin{align*}
D_{t} =U_{t} +\epsilon_{t},
\end{align*}
where $\epsilon_{t} \sim \N(0,R)$ for $t\in [0,\tau ]$. Here $U_{t}$ represents the underlying spatio-temporal degradation process and $\epsilon_{t}$ is the noise process. We use the variance-covariance matrix $R$ to describe the spatial correlation among the noise terms at different locations.

The modeling of $U_t$ is typically more challenging than the traditional type of degradation data. It is of benefit to leverage physical knowledge about the process. The degradation process usually results from a physical process, which can be modeled by a system of partial differential equations (e.g., describing a diffusion process) with a set of boundary conditions. Using finite differences to approximate the partial derivatives and adding a random term to make it stochastic, we have the following model for the true degradation process,
\begin{align*}
U_{t}=&MU_{t-1} +M^{(b)}U^{(b)}_{t-1}+\eta_{t},\\
U_{0}\sim & N(\mu_{0}, \Sigma_{0}).
\end{align*}
Here $\eta_{t}$ is the temporal random effect that follows a $\N(0,Q)$ distribution with variance-covariance matrix $Q$, $M$ is the propagator matrix, $M^{(b)}$ is the boundary propagator matrix, and $U^{(b)}_{t}$ is the boundary condition. The initial status for the process is given by $U_{0}$ having a multivariate normal distribution with mean $\mu_{0}$ and variance-covariance matrix $\Sigma_{0}$.

Parameter estimation can be done by using Bayesian methods, which requires the specification of prior information. For example, we can use inverse Wishart distributions to specify the prior for $R$ and $Q$, and a Gaussian process prior for the propagator matrices. We can either set the hyperparameters to be fixed or assign prior distributions to these parameters as well. Then the posterior distribution for $U_{t}, t\in [0,\tau]$, is
\begin{align*}
\pi_{(U_{0} ,U_{1}, \cdots, U_{\tau}, M, Q, R |D_{1},\cdots, D_{\tau})}\propto & \left[\prod_{t=1}^{\tau }\pi_{(D_{t} | U_{t,} R)}\right]\times\pi_{U_{0}}\\
&\times\left[\prod_{t=1}^{\tau }\pi_{(U_{t} | U_{t-1,} M,Q)}\right]\times\pi_{R}\times\pi_{Q}\times\pi_{M}.
\end{align*}
Here $\pi_{(\cdot|\cdot)}$ and $\pi_{(\cdot)}$ on the right hand side of the equation denote conditional distributions and prior distributions, respectively. The posterior distribution can be obtained by using Markov chain Monte Carlo (MCMC) via a Gibbs sampler. Based on the estimated model and failure definition, one can make reliability predictions using MCMC methods. A failure event can be defined when the maximum of the degradation measurement exceeds a threshold, or when area of the region that is above a certain degradation level exceeds a threshold.

\section{Reliability Models using Covariates}\label{sec:tackling.covariates}
In traditional reliability field data modeling and analysis, covariate information (when is available) tends to be simple such as operating temperature and other conditions, while the response used is typically time to event, degradation levels, or recurrent event times.
Recent development of reliability methods include \citeN{ZhuYashchinHosking2014}, \citeN{Peng2016}, and \shortciteN{Wangetal2017}. Regarding covariate modeling, the system operating and environmental (SOE) data have become available for many different kinds of systems. The SOE data essentially are multivariate time-varying covariates. With different types of responses for reliability models, the SOE data can be integrated to provide better reliability predictions. Recent developments include integrating SOE data with time to event data (e.g., \citeNP{HongMeeker2013}, \shortciteNP{Yokoyama2015}, \shortciteNP{ChiharuKumazaki2015}, and \citeNP{MasahiroYokoyam2016}), with degradation data (e.g., \shortciteNP{HongDuanetal2015}, and \shortciteNP{XuHongJin2015}), and with recurrent event data (e.g., \shortciteNP{Xuetal2017}). This section describes some other directions for using covariates in reliability data modeling and analysis.

\subsection{Variable Selection}
Although the SOE data contain large amounts of information, it is not necessarily the case that all information is relevant for the event process or the degradation process. One needs to reduce the large $p$ number of covariates and only keep relevant covariates for a parsimonious model. Here we describe some general ideas of an automatic method for variable selection. In statistics literature, there have been many recent developments in variable selection, especially via penalty functions. Thus, the idea of using penalty functions in reliability setting will be discussed here.

Let $\mathcal{L}_i(\thetavec; w_i)$ be the likelihood contribution of the data from unit $i$ based on a reliability model (e.g., a time to event model or a degradation model) that incorporates the SOE covariate information. Here $\thetavec$ is a general parameter vector of the unknown model parameters. The model typically involves random effects $w_i$ which are unobservable. The penalized maximum likelihood (ML) approach will be used for parameter estimation. The penalized log likelihood is defined as
\begin{align}\label{eqn:likelihood.fun}
l(\thetavec)=&-\sum_{i=1}^{n}\log\left[\int_{w_i}\mathcal{L}_i(\thetavec; w_i)f_{\nor}(w_i)dw_i\right]+P_{\alphavec}(\betavec),
\end{align}
where $f_{\nor}(w_i;\sigma_w)$ is the probability density function (pdf) of a normal distribution, $\betavec=(\beta_1,\cdots, \beta_p)'$ are the coefficients for the covariates, and $P_{\alphavec}(\betavec)$ is a penalty function on $\betavec$ with penalty $\alphavec$. The penalty function $P_{\alpha}(\betavec)$ in \eqref{eqn:likelihood.fun} will allow keeping the subset of the covariates that are most important for describing the event process or the degradation process.

For the SOE data, the use of an elastic net (EN) penalty function in \citeN{ZouHastie2005} is suitable. The EN penalty function can be expressed as $P_{\alphavec}(\betavec)=\alpha_1\sum_{l=1}^p|\beta_l|+\alpha_2\sum_{l=1}^p\beta_l^2$, where $\alphavec=(\alpha_1,\alpha_2)'$ are the penalty values. The EN penalty is a combination of an $L_1$ penalty and an $L_2$ penalty. The first term $\alpha_1\sum_{l=1}^p|\beta_l|$ is the LASSO penalty, which is the most commonly-used penalty function (e.g., \citeNP{HastieTibshiraniFriedman2009}). The second term $\alpha_2\sum_{l=1}^p\beta_l^2$ is the ridge-regression type penalty (e.g., \citeNP{HoerlKennard1970}). The LASSO penalty will automatically set the coefficients of non-important variables to zero to achieve sparsity. When there is a group of variables among which the pairwise correlations are high, the LASSO penalty, when used by itself, tends to select only one variable from the group and does not care which one is selected (\citeNP{ZouHastie2005}). The SOE data contain many variables and usually there are high correlations between pairs of variables. The $L_2$~penalty tends to work better in situations with a high degree of collinearity but it keeps all covariates in the model. The EN penalty has the advantage of using both of the $L_1$ and $L_2$ penalties.

\subsection{Functional Covariates}
\begin{figure}
\begin{center}
\begin{tabular}{cc}
\includegraphics[width=0.45\textwidth]{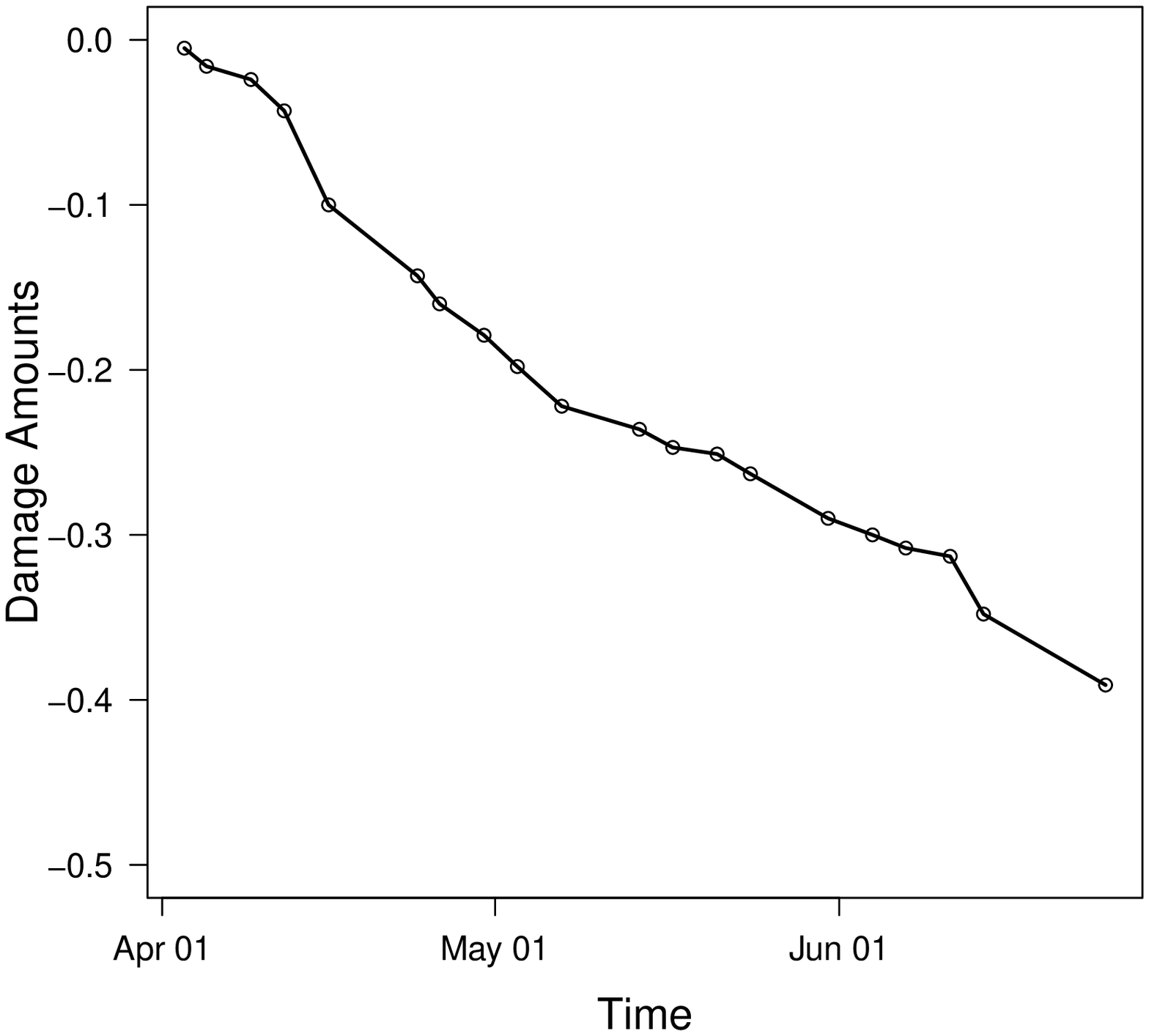}&
\includegraphics[width=0.45\textwidth]{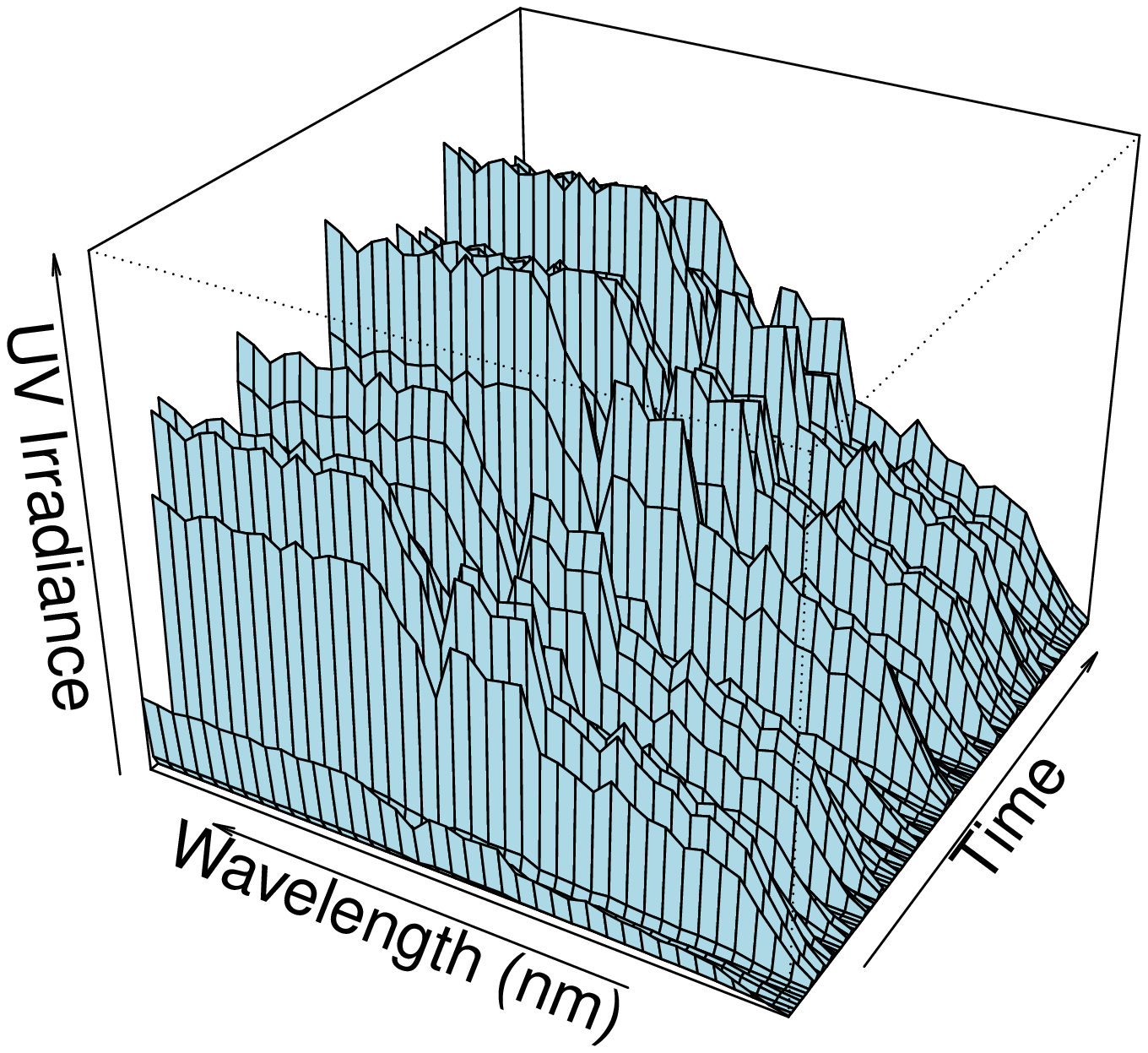}\\
(a) Degradation  & (b) UV information
\end{tabular}
\caption{Example of degradation data for one unit (a) and functional covariates showing the UV intensity (b) from NIST coating data.}\label{fig:coating.uv}
\end{center}
\end{figure}

In addition to the traditional covariates (time invariant and time varying), some covariates take the form of a functional curve. Functional covariates sometimes arise in reliability analysis. That is, the response in the dataset is the time to event or degradation measurement, while the covariates are functional curves that characterize certain features of the product and system that can be related to the product reliability. Figure~\ref{fig:coating.uv}(a) shows an example degradation path from the NIST coating data and Figure~\ref{fig:coating.uv}(b) shows the UV profile that is related to the degradation data (along with temperature and relative humidity). The UV information at a specific point is a functional curve of UV wavelength $\lambda$, as shown in Figure~\ref{fig:coating.uv}(b). Although the coating data has been analyzed in \shortciteN{HongDuanetal2015}, the UV information was simply aggregated over the range of the wavelength. That is, a scalar is used to represent the functional covariate at each time point.

With function data analysis techniques, the UV information can be treated as a functional covariate. One can flexibly model the effect of UV wavelength $\lambda$. Let $y_{ij}$ be the degradation measurement at $t_{ij}$ for unit $i$. Also, let $x_{i}(\lambda; t)$ be the UV intensity at time $t$ and wavelength $\lambda$  for unit $i$. The following functional regression framework can be considered. That is
\begin{align}\label{eqn:fda.covariate}
y_{ij}=\beta_0+\sum_{t=0}^{t_{ij}}\int_{\lambda}\psi(\lambda)x_{i}(\lambda; t)d\lambda+\epsilon_{ij}.
\end{align}
Here $\psi(\lambda)$ describes the functional effect that UV wavelength has on degradation $y_{ij}$. Due to the monotonicity of the degradation path, we use a cumulative damage model here which is represented by the summation over time $t$ in \eqref{eqn:fda.covariate}. In the regression framework, the ordinary time-varying covariates $x_i(t)$ (e.g., temperature and relative humidity) can be easily incorporated by adding an extra term $\sum_{t=0}^{t_{ij}}\beta x_{i}(t)$ into \eqref{eqn:fda.covariate}. Spline bases can be used to construct $\psi(\lambda)$ so that a flexible functional form can be achieved for the effect of UV wavelength.

\subsection{Image Covariates}
\begin{figure}
\begin{center}
\begin{tabular}{cc}
\includegraphics[width=0.3\textwidth]{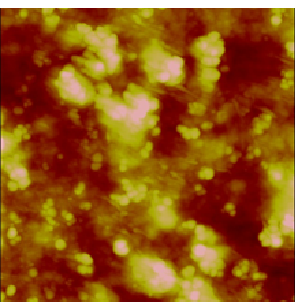}&
\includegraphics[width=0.3\textwidth]{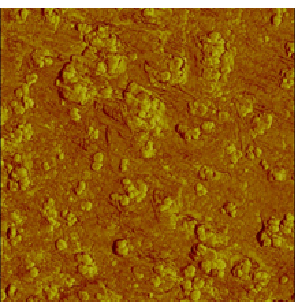}\\
(a) Amplitude & (b) Phase
\end{tabular}
\caption{Illustration of image covariates from AFM images of a material.}\label{fig:micro.structure}
\end{center}
\end{figure}

In some applications, the response is the time to event or a degradation measurement, and images are used to characterize a property of the unit. For example, \citeN{SiYangWu2016} studied how to use images of the microstructure of a high strength steel to predict the lifetime of the material. For another example, Figure~\ref{fig:micro.structure} shows an example of atomic force microscopy (AFM) images for the microstructure of a polymer material tested at NIST. In this case, the response might be the breaking strength of the material. Thus, it is practically useful if one can build a model to predict the strength to fail as a function of the microstructure images.

In statistics, such problems have been studied as an image regression problem. The tensor regressor models in \citeN{ZhouLiZhu2013} provide a general way to incorporate images as covariates. In particular, a tensor is defined as a multi-dimensional array, in which a two-dimension image $X$ is a special case of a tensor. \citeN{ZhouLiZhu2013} proposed a general linear model framework for image regression, in which a link function $g(\cdot)$ is used to link the mean of the response denoted by $\mu$ to the image covariate. That is
\begin{align}\label{eqn:tensor.reg}
g(\mu)=\alpha+\langle B, X\rangle,
\end{align}
where $\alpha$ is a constant term and, $\langle\cdot,\cdot\rangle$ is the inner product of two tensors. Here $B$ is a tensor containing the regression parameters with certain structures so that its elements can be estimated. The details of the estimation can be found in \citeN{ZhouLiZhu2013}. \citeN{WangZhu2017} proposed a scalar-on-image regression framework. While the general model is the same as in \eqref{eqn:tensor.reg}, \citeN{WangZhu2017} modeled the tensor coefficient $B$ as a piecewise smooth function and developed a new estimation method based on total variation analysis. The challenge of applying such advanced methods to lifetime regression is the need to deal with censoring. For degradation data with image covariates, handling correlations among images taken at different time points is also a challenging problem. However, these challenges provide opportunities for future research.

\section{Machine Learning and Reliability Analysis}\label{sec:machine.learning.rel.pred}
Machine learning can provide useful tools for reliability analysis, especially in dealing with data complexity. Here we discuss several machine learning techniques that can be used for reliability analysis.

\subsection{Signal Clustering}
\begin{figure}
\begin{center}
\begin{tabular}{cc}
\includegraphics[width=0.45\textwidth]{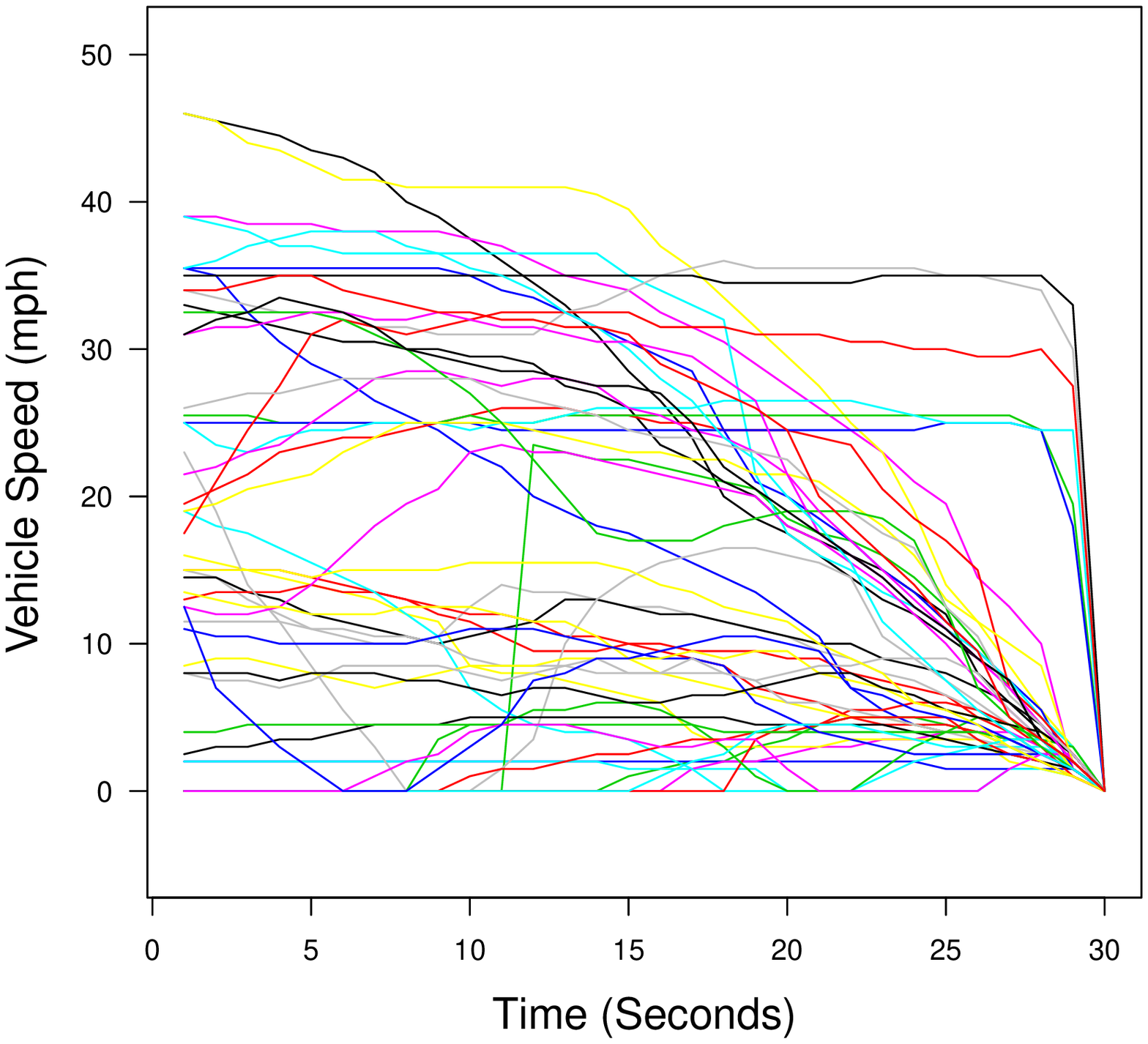}&
\includegraphics[width=0.45\textwidth]{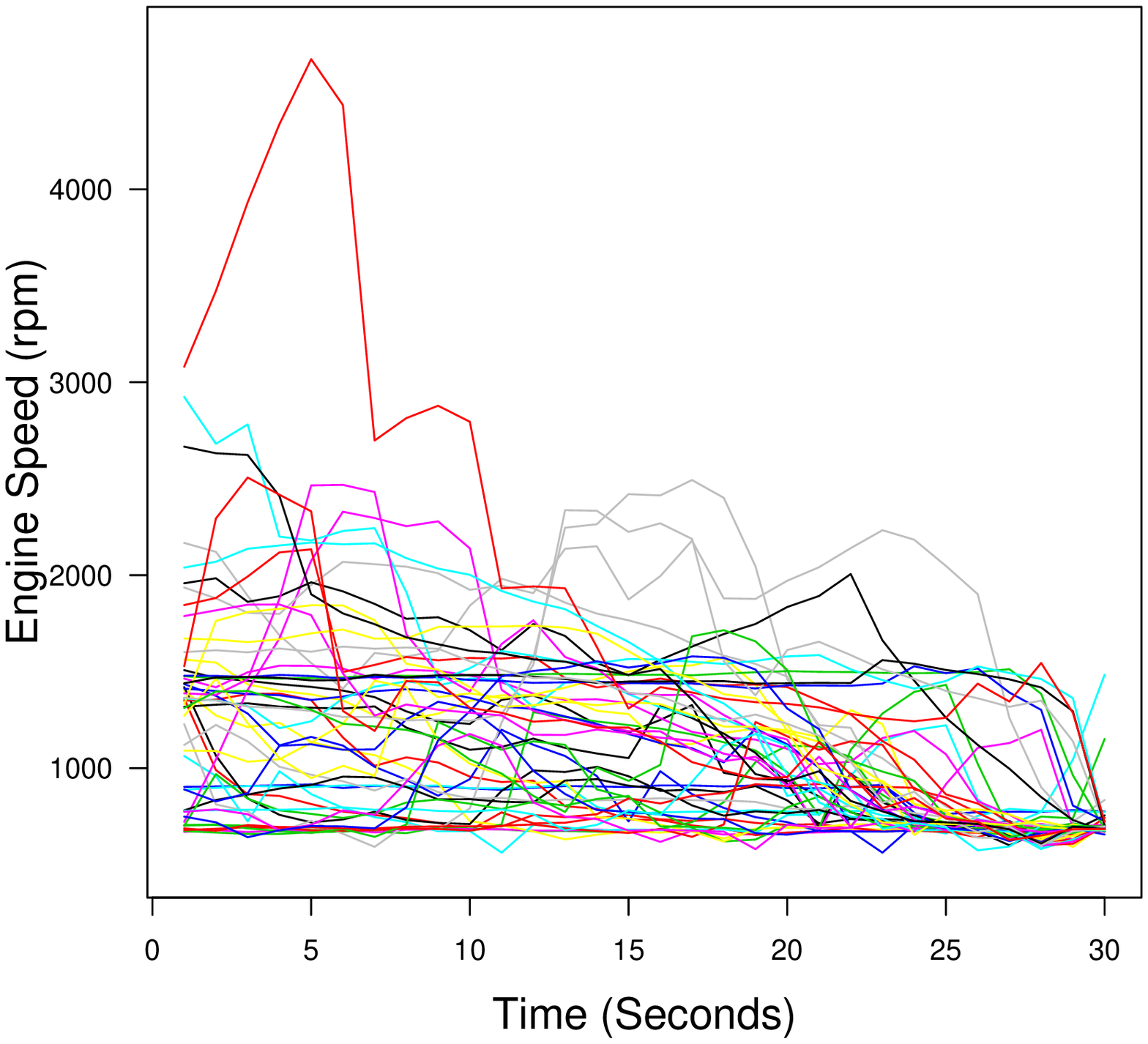}\\
(a) Variable A  & (b) Variable B \\
\end{tabular}
\caption{Example of sensor data for two channels with respect to specifically defined events.}\label{fig:sesor.data.clustering}
\end{center}
\end{figure}

With the increasing development of sensor and communication technologies, multi-channel sensor signals can be obtained in real time arriving at a high rate (e.g., per second or millisecond). Hence, data generated by these sensors are increasing enormously due to the high volume. From online signals, certain events can be detected, and it is often of interest to categorize those events into different types. Thus, different actions can be taken for different types of events. These sensor data can be treated as multivariate functional data. Because not all variables are useful for clustering, it is important to include only those variables that are useful in the clustering procedure. Figure~\ref{fig:sesor.data.clustering} shows an example of sensor data for two channels with respect to specific defined events. In the literature, \citeN{JacquesPreda2014} considered multivariate functional data clustering. \citeN{WangZhu2008} studied model based clustering with variable selections. Here we briefly describe a clustering algorithm for multivariate functional data with automatic variable selection.

We use functional principal component analysis to transform the functional data into multivariate data. The multivariate functional data can be decomposed according to the Karhunen-Loeve expansion. In this setting, the multivariate functional data $X(t)$ can be represented as
\begin{equation} \label{eqn:decomposition}
X(t)=\mu(t)+\sum_{l=1}^{\infty }\sum_{j=1}^{p} C_{lj} f_{lj}(t) ,\quad t\in [0,T].
\end{equation}
Here, $\mu(t)$ is the mean function, $f_{lj}(t)$ are functional principal factors, and $C_{lj}$ are principal components. Similar to univariate cases, $C_{lj} $ are uncorrelated random variables with mean zero and variance $\lambda_{lj} $. In practice, the expansion in \eqref{eqn:decomposition} is often truncated to provide the series with a finite number of terms. That is,
$$X(t)\approx \mu(t)+\sum_{l=1}^{L}\sum_{j=1}^{p} C_{lj} f_{lj}(t),\quad t\in [0,T].$$
Here, $L$ should be chosen to be large enough so that at least 95\% of the variation in the data is explained. Using the available data, $X_{i}$, the principal components $C_{lj}$ can be estimated using functional data analysis techniques, which are denoted by $c_{ilj}$. In this way, the functional data $X_{i}$ are transformed to multivariate data $c_{ilj}$.

We then use a model-based algorithm with a penalty term to do clustering. The method does clustering with automatic variable selection through a penalty term, assuming the data are generated from a Gaussian mixture distribution. That is, the pdf is
$$g(c_{i})=\sum_{k=1}^{K} \pi_{k} g_{k}(c_{i};\mu _{k},\Sigma),$$
where $c_{i}$ is a vector containing the $c_{ilj}$ values with the same $i$, and $g_{k}(\cdot)$ is a pdf with mean $\mu_{k}$ and covariance matrix $\Sigma$.  Here the mean vector is $\mu _{k}$ containing $\mu _{klj}$ values and the coordinate of variable $(lj)$ corresponding to cluster $k$. The constant $K$ is a pre-set value for the number of clusters (usually it is large enough). The algorithm will automatically select the number of clusters based on the data. Let $\delta_{ik} $ be the indicator for membership of unit $i$ in cluster $k$. The log likelihood function with a penalty term is
$$l_{P}(\theta)=\sum_{i=1}^{n}\sum_{k=1}^{K} \delta_{ik}[\log\pi_{k} +\log f_{k}(c_{i}; \mu_{k}, \Sigma)]-\lambda \sum_{j=1}^{p} \max_{k,l} (|\mu_{klj} |).$$
The penalty $\lambda \sum_{j=1}^{p} \max_{k,l} (|\mu_{klj}|)$ will automatically remove those variables with little contribution to the clustering. Because the $\delta_{ik}$ indicator parameters are not observed, the expectation and maximization (EM) algorithm can be used for parameter estimation.

\subsection{Deep Learning and Reliability Prediction}
Deep learning methods were developed to solve prediction problems. Because many problems in reliability involve predictions, there seems to be a natural intersection between these two areas. However, we see little applications of deep learning in reliability. Traditionally, reliability predictions are mostly based on parametric models due to the need for extrapolation. The arrival of big data provides the opportunity to apply deep learning method in reliability predictions. A good resource for deep learning is \citeN{Goodfellow-et-al-2016}.

The complex nature of big data provides different types of covariates, and forms complicated covariate history. Product usage history becomes so complicated that it will be a challenge to use covariate within the traditional regression framework. One idea is to make reliability predictions based on similarity of product usage histories. With multi-layers of a deep network, the method can automatically match individual histories and thus be expected to provide accurate predictions when failures are driven by usage variables.

\subsection{Text Analytics}
Text data also appears in reliability datasets, especially in warranty claim databases, maintenance databases, and after service businesses. For example, customer complaints may indicate product reliability issues. Also, technicians comments about the diagnostics and repairs provide failure information. With the wide availability of smartphones, online reviews, and social media, text becomes an important format of unstructured data for extracting reliability information. Text analytics and text mining are important tools for achieving that goal.

\citeN{DeovratKakde2015} discussed leveraging unstructured data to detect emerging reliability issues. Customer complaint texts were used for text mining to identify useful topics or customer concerns. For example, customers may discuss topics such as airbags, engine, fuel system, and power steering locks. If over a period of time there is a detection of an increase of a particular topic, such as power steering locks, this may provide an indication that the power steering system may have emerging reliability issues.

\section{Emerging Application Areas}\label{sec:new.areas}
 \subsection{Weathering and Usage-Based Warranty Predictions}
Big data will bring important changes to the practice of warranty prediction and spare parts provisioning.  \citeN{XiaoLiu2015} presented a description of reliability analysis and spares provisioning for repairable systems with dependent failure processes and a time-varying installed base. \shortciteN{ShahLimon2016} presented reliability estimation considering use rate profiles and warranty claims. \citeN{KingHongMeeker2017} used product component genealogy information, which is readily available in production databases, to improve prediction accuracy. \shortciteN{Heetal2018} developed a warranty prediction method that uses a concept of learning effects in reliability and used a log-linear regression model to estimate learning effects.

With the advancement of reliability analysis techniques and big data, it is possible to generate individualized, not-one-size-fit-all, warranty predictions. Certain products such as a coating that are sensitive weather conditions can have personalized prediction using the spatio-temporal covariates about the weather condition in the products environment. Products subject to different use profiles, such as automobiles, can also have different warranty predictions based on the usage information.

\subsection{Early Warning Systems for Reliability and Warranty Data}
Early warning of reliability issues of systems and products is important for many applications. Such early warning can be issued based on monitoring of claims in warranty database and reliability data from the field. \citeN{WuMeeker2002} developed an early detection procedure for reliability problems using information from warranty databases, which is a nonparametric approach based on claim counts. \citeN{Yashchin2012} used a CUSUM (i.e., cumulative sum control chart) based approach to monitor lifetime data streams, which can detect abrupt changes in parameters that could be related to reliability issue. The developed procedure was deployed in the IBM Personal Systems Division. \citeN{lawless2012monitoring} developed CUSUM procedures for monitoring warranty claims, which can detect changes in claim rates in a timely manner. With the help of big data, especially the real-time availability of data streams that contain reliability information, more advanced procedures can be developed for early warning systems for reliability and warranty data.

\subsection{Usage-Based Insurance}
Usage based insurance (UBI) is a type of insurance in which the cost of the insurance is based on the usage, use behavior, and product location. One important task is to access the future risk at individual level to develop pricing plans. With some modifications and customization, reliability models that are used to assess product failure risks can be used to predict events for UBI. Thus, UBI will be an interesting area to apply reliability analysis techniques.

\subsection{Energy Sector}
Due to the development of renewable energy technology, the reliability of photovoltaic (PV) systems and components is great interest. One of the goals in PV industry is to demonstrate that the PV products can last at least 30 years so that banks will be willing to finance investments of large-scale PV systems. Due to highly compact and integrated nature of PV modules, assessing field reliability via laboratory accelerated tests is challenging. Also, PV systems have multiple failure mechanisms such as material degradation as well as mechanical and electrical component failures. This complexity provides opportunities to develop new reliability analysis methods for test planning and reliability prediction.

A smart grid is an extremely complicated system which connects power generation, transmission, and distribution systems, involving substation, factories, home and building, and even with those distributed generation systems. Reliability is important for smart-grid system maintenance. The data collected from a smart grid tend to have a large volume but also with a high level of variety and complexity. Traditional reliability analysis methods need to be extended, which opens doors for research to develop new reliability analysis methods.

\subsection{Internet of Things}
The Internet of Things (IoT) is a broadly defined network in which things and objects can all be connected. The introduction of such connectors has started to change people's life. Conceptually, the IoT has four layers: an application layer, a connectivity layer, an information processing layer, and a physical layer. The IoT, like other products or systems, will also fail during operation. Thus, reliability problems also exist in the IoT. Reports of current research on the reliability of the IoT systems is sparse. \citeN{ZinTinHama2016} discussed the reliability and availability measures for the IoT. Thus, the IoT will become an emerging area for reliability analysis, including methods to define reliability metrics, reliability tests, data collection, and the development of methods to make reliability predictions.

\subsection{Big Data and Test Planning}
Reliability test planning has been a challenging problem due to limited time, budget, and facility, and the fact that products may last a long time in the field (e.g., \shortciteNP{Hongetal2015}). How big data technology can be applied in the design of experiments for reliability study is an interesting area of research. For example, \shortciteN{Kingetal2016} studied the test planning of polymer composites under an ordinary test planning framework, where the response was the cycles to failure and the accelerating variable was the level of stress applied to the test coupon. With advanced measurement instruments, it is possible to measure how microstructural changes occur during the accelerated test process, which can possibly lead to a better understanding of the failure mechanism and prediction of lifetime at use conditions. With the help of physical models and physics-based computer simulation models, it is possible to use microstructure data to build reliability models with stronger predictive power, which can lead to a reduction in test duration and the number of samples that need to be tested.

\section{Concluding Remarks}\label{sec:conclusions}
In this paper, we provide a review and discussion on some aspects of reliability analysis that is related to the complexity dimension of big data. In summary, the arrival of big data provides lots of opportunities for developing and applying new statistical methods. \citeN{Anderson-Cook2015} provided a comprehensive discussion on opportunities for statistical research in emerging areas. In reliability analysis, the complexity aspect of big data provides lots of opportunities for applying existing methods and expansion of areas of reliability analysis.

We observe the following trend. New technology leads to new types of data, and eventually leads to new statistical methods. This trend has been observed in reliability and the development of new statistical techniques such as functional data analysis, and image regression provide new tools for reliability analysis. Reliability analysis techniques can also be applied to new areas such as renewable energy and to solve important emerging problems.

Our discussion focuses mostly on data modeling and analysis for reliability prediction. Big data also brings many opportunities in other areas of reliability such as maintenance. For example, \citeN{Zhang2015} discussed big data and its applications in maintenance, especially through conditional monitoring and fault detection. Big data can be expected to become a major driving force for the innovation on reliability data analysis and reliability engineering in the broader picture.

\section*{Acknowledgments}
The authors thank the editors for their valuable comments that helped in improving this paper.  The authors acknowledge Advanced Research Computing at Virginia Tech for providing computational resources. The work by Hong was partially supported by the National Science Foundation under Grants CMMI-1634867 and CNS-1565314 to Virginia Tech.


\begin{thebibliography}{}

\bibitem[\protect\citeauthoryear{Anderson-Cook}{Anderson-Cook}{2015}]{Anderson-Cook2015}
Anderson-Cook, C.~M. (2015).
\newblock Opportunities to empower statisticians in emerging areas.
\newblock {\em Applied Stochastic Models in Business and Industry\/}~{\em 31},
  3--11.

\bibitem[\protect\citeauthoryear{Bedair, Hong, Li, and Al-Khalidi}{Bedair
  et~al.}{2016}]{Bedairetal2016}
Bedair, K., Y.~Hong, J.~Li, and H.~R. Al-Khalidi (2016).
\newblock Multivariate frailty models for multi-type recurrent event data and
  an application to cancer prevention trial.
\newblock {\em Computational Statistics and Data Analysis\/}~{\em 101},
  161--173.

\bibitem[\protect\citeauthoryear{Chehade, Song, Liu, Saxena, and Zhang}{Chehade
  et~al.}{2018}]{Chehadeetal2018}
Chehade, A., C.~Song, K.~Liu, A.~Saxena, and X.~Zhang (2018).
\newblock A data-level fusion approach for degradation modeling and prognostic
  analysis under multiple failure modes.
\newblock {\em Journal of Quality Technology\/}~{\em 50}, xx--xx.

\bibitem[\protect\citeauthoryear{Chen and Ye}{Chen and Ye}{2018}]{ChenYe2018}
Chen, P. and Z.~Ye (2018).
\newblock Uncertainty quantification for monotone stochastic degradation
  models.
\newblock {\em Journal of Quality Technology\/}~{\em 50}, xx--xx.

\bibitem[\protect\citeauthoryear{Cressie and Wikle}{Cressie and
  Wikle}{2011}]{CressieWikle2011}
Cressie, N. and C.~K. Wikle (2011).
\newblock {\em Statistics for Spatio-Temporal Data}.
\newblock Hoboken, NJ: John Wiley \& Sons.

\bibitem[\protect\citeauthoryear{Escobar, Meeker, Kugler, and Kramer}{Escobar
  et~al.}{2003}]{EscobarMeekerKuglerKramer2003}
Escobar, L.~A., W.~Q. Meeker, D.~L. Kugler, and L.~L. Kramer (2003).
\newblock Accelerated destructive degradation tests: Data, models, and
  analysis.
\newblock In B.~H. Lindqvist and K.~A. Doksum (Eds.), {\em Mathematical and
  Statistical Methods in Reliability}. Singapore: World Scientific Publishing
  Company.

\bibitem[\protect\citeauthoryear{Fang, Paynabar, and Gebraeel}{Fang
  et~al.}{2017}]{FangPaynabarGebraeel2017}
Fang, X., K.~Paynabar, and N.~Gebraeel (2017).
\newblock Multistream sensor fusion-based prognostics model for systems with
  single failure modes.
\newblock {\em Reliability Engineering \& System Safety\/}~{\em 159}, 322--331.

\bibitem[\protect\citeauthoryear{Gao and Meeker}{Gao and
  Meeker}{2012}]{GaoMeeker2012}
Gao, C. and W.~Q. Meeker (2012).
\newblock A statistical method for crack detection from vibrothermography
  inspection data.
\newblock {\em Quality Technology \& Quantitative Management\/}~{\em 9},
  59--77.

\bibitem[\protect\citeauthoryear{Goodfellow, Bengio, and Courville}{Goodfellow
  et~al.}{2016}]{Goodfellow-et-al-2016}
Goodfellow, I., Y.~Bengio, and A.~Courville (2016).
\newblock {\em Deep Learning}.
\newblock MIT Press.

\bibitem[\protect\citeauthoryear{Gu, Dickens, Stanley, Byrd, Nguyen,
  Vaca-Trigo, Meeker, Chin, and Martin}{Gu et~al.}{2009}]{Guetal2008}
Gu, X., B.~Dickens, D.~Stanley, W.~E. Byrd, T.~Nguyen, I.~Vaca-Trigo, W.~Q.
  Meeker, J.~W. Chin, and J.~W. Martin (2009).
\newblock Linking accelerating laboratory test with outdoor performance results
  for a model epoxy coating system.
\newblock In J.~Martin, R.~A. Ryntz, J.~Chin, and R.~A. Dickie (Eds.), {\em
  Service Life Prediction of Polymeric Materials}. NY: New York: Springer.

\bibitem[\protect\citeauthoryear{Hastie, Tibshirani, and Friedman}{Hastie
  et~al.}{2009}]{HastieTibshiraniFriedman2009}
Hastie, T., R.~Tibshirani, and J.~Friedman (2009).
\newblock {\em The Elements of Statistical Learning: Data Mining, Inference,
  and Prediction\/} (Second ed.).
\newblock Springer.

\bibitem[\protect\citeauthoryear{He, Zhang, Jiang, and Bian}{He
  et~al.}{2018}]{Heetal2018}
He, S., Z.~Zhang, W.~Jiang, and D.~Bian (2018).
\newblock Predicting field reliability based on two-dimensional warranty data
  with learning effects.
\newblock {\em Journal of Quality Technology\/}~{\em 50}, xx--xx.

\bibitem[\protect\citeauthoryear{Hoerl and Kennard}{Hoerl and
  Kennard}{1970}]{HoerlKennard1970}
Hoerl, A.~E. and R.~W. Kennard (1970).
\newblock Ridge regression: Biased estimation for nonorthogonal problems.
\newblock {\em Technometrics\/}~{\em 12}, 55--67.

\bibitem[\protect\citeauthoryear{Hong, Duan, Meeker, Stanley, and Gu}{Hong
  et~al.}{2015}]{HongDuanetal2015}
Hong, Y., Y.~Duan, W.~Q. Meeker, D.~L. Stanley, and X.~Gu (2015).
\newblock Statistical methods for degradation data with dynamic covariates
  information and an application to outdoor weathering data.
\newblock {\em Technometrics\/}~{\em 57}, 180--193.

\bibitem[\protect\citeauthoryear{Hong, King, Zhang, and Meeker}{Hong
  et~al.}{2015}]{Hongetal2015}
Hong, Y., C.~B. King, Y.~Zhang, and W.~Q. Meeker (2015).
\newblock Bayesian life test planning for log-location-scale family of
  distributions.
\newblock {\em Journal of Quality Technology\/}~{\em 47}, 336--350.

\bibitem[\protect\citeauthoryear{Hong and Meeker}{Hong and
  Meeker}{2013}]{HongMeeker2013}
Hong, Y. and W.~Q. Meeker (2013).
\newblock Field-failure predictions based on failure-time data with dynamic
  covariate information.
\newblock {\em Technometrics\/}~{\em 55}, 135--149.

\bibitem[\protect\citeauthoryear{Jacques and Preda}{Jacques and
  Preda}{2014}]{JacquesPreda2014}
Jacques, J. and C.~Preda (2014).
\newblock Model-based clustering for multivariate functional data.
\newblock {\em Computational Statistics and Data Analysis\/}~{\em 71}, 92--106.

\bibitem[\protect\citeauthoryear{Kakde and Chaudhuri}{Kakde and
  Chaudhuri}{2015}]{DeovratKakde2015}
Kakde, D. and A.~Chaudhuri (2015).
\newblock Leveraging unstructured data to detect emerging reliability issues.
\newblock {\em Reliability and Maintainability Symposium (RAMS), 2015
  Proceedings - Annual\/}, 1--5.

\bibitem[\protect\citeauthoryear{King, Hong, Dehart, Defeo, and Pan}{King
  et~al.}{2016}]{Kingetal2016}
King, C., Y.~Hong, S.~P. Dehart, P.~A. Defeo, and R.~Pan (2016).
\newblock Planning fatigue tests for polymer composites.
\newblock {\em Journal of Quality Technology\/}~{\em 48}, 227--245.

\bibitem[\protect\citeauthoryear{King, Hong, and Meeker}{King
  et~al.}{2017}]{KingHongMeeker2017}
King, C., Y.~Hong, and W.~Q. Meeker (2017).
\newblock Product component genealogy modeling and field-failure prediction.
\newblock {\em Quality and Reliability Engineering International\/}~{\em 33},
  135--148.

\bibitem[\protect\citeauthoryear{Kumazaki, Yamamoto, and Suzuki}{Kumazaki
  et~al.}{2015}]{ChiharuKumazaki2015}
Kumazaki, C., W.~Yamamoto, and K.~Suzuki (2015).
\newblock Lifetime prediction of vehicle components using online monitoring
  data.
\newblock {\em Total Quality Science\/}~{\em 1}, 52--64.

\bibitem[\protect\citeauthoryear{Lawless, Crowder, and Lee}{Lawless
  et~al.}{2012}]{lawless2012monitoring}
Lawless, J.~F., M.~J. Crowder, and K.-A. Lee (2012).
\newblock Monitoring warranty claims with cusums.
\newblock {\em Technometrics\/}~{\em 54}, 269--278.

\bibitem[\protect\citeauthoryear{Limon, Yadav, Zuo, Muscha, and Honeyman}{Limon
  et~al.}{2016}]{ShahLimon2016}
Limon, S., O.~P. Yadav, M.~J. Zuo, J.~Muscha, and R.~Honeyman (2016).
\newblock Reliability estimation considering usage rate profile and warranty
  claims.
\newblock {\em Proceedings of the Institution of Mechanical Engineers, Part O:
  Journal of Risk and Reliability\/}~{\em 230}, 297--308.

\bibitem[\protect\citeauthoryear{Liu, Gebraeel, and Shi}{Liu
  et~al.}{2013}]{LiuGebraeelShi2013}
Liu, K., N.~Gebraeel, and J.~Shi (2013).
\newblock A data-level fusion model for developing composite health indices for
  degradation modeling and prognostic analysis.
\newblock {\em IEEE Transactions on Automation Science and Engineering\/}~{\em
  10}, 652--664.

\bibitem[\protect\citeauthoryear{Liu and Tang}{Liu and
  Tang}{2015}]{XiaoLiu2015}
Liu, X. and L.~C. Tang (2015).
\newblock Reliability analysis and spares provisioning for repairable systems
  with dependent failure processes and a time-varying installed base.
\newblock {\em IIE Transactions\/}~{\em 48}, 43--56.

\bibitem[\protect\citeauthoryear{Liu, Yeo, and Kalagnanam}{Liu
  et~al.}{2018}]{LiuYeoKalagnanam2016}
Liu, X., K.~Yeo, and J.~Kalagnanam (2018).
\newblock A statistical modeling approach for spatio-temporal degradation data.
\newblock {\em Journal of Quality Technology\/}~{\em 50}, xx--xx.

\bibitem[\protect\citeauthoryear{Lu and Meeker}{Lu and
  Meeker}{1993}]{LuMeeker1993}
Lu, C.~J. and W.~Q. Meeker (1993).
\newblock Using degradation measures to estimate a time-to-failure
  distribution.
\newblock {\em Technometrics\/}~{\em 34}, 161--174.

\bibitem[\protect\citeauthoryear{Meeker and Escobar}{Meeker and
  Escobar}{1998}]{meekerescobar1998}
Meeker, W.~Q. and L.~A. Escobar (1998).
\newblock {\em Statistical Methods for Reliability Data}.
\newblock New York: John Wiley \& Sons, Inc.

\bibitem[\protect\citeauthoryear{Meeker and Hong}{Meeker and
  Hong}{2014}]{MeekerHong2014}
Meeker, W.~Q. and Y.~Hong (2014).
\newblock Reliability meets big data: Opportunities and challenges, with
  discussion.
\newblock {\em Quality Engineering\/}~{\em 26}, 102--116.

\bibitem[\protect\citeauthoryear{Meyer}{Meyer}{2008}]{Meyer2008}
Meyer, M.~C. (2008).
\newblock Inference using shape-restricted regression splines.
\newblock {\em The Annals of Applied Statistics\/}~{\em 2}, 1013--1033.

\bibitem[\protect\citeauthoryear{Nelsen}{Nelsen}{2006}]{Nelsen2006}
Nelsen, R.~B. (2006).
\newblock {\em An {Introduction} to {Copulas}\/} (second ed.).
\newblock New York: Springer.

\bibitem[\protect\citeauthoryear{Nelson}{Nelson}{1990}]{Nelson1990}
Nelson, W. (1990).
\newblock {\em Accelerated Testing: Statistical Models, Test Plans, and Data
  Analyses, (Republished in a paperback in Wiley Series in Probability and
  Statistics, 2004)}.
\newblock New York: John Wiley \& Sons.

\bibitem[\protect\citeauthoryear{Pan and Balakrishnan}{Pan and
  Balakrishnan}{2011}]{PanBalakrishnan2011}
Pan, Z. and N.~Balakrishnan (2011).
\newblock Reliability modeling of degradation of products with multiple
  performance characteristics based on gamma processes.
\newblock {\em Reliability Engineering \& System Safety\/}~{\em 96}, 949--957.

\bibitem[\protect\citeauthoryear{Pan, Balakrishnan, Sun, and Zhou}{Pan
  et~al.}{2013}]{Panetal2013}
Pan, Z., N.~Balakrishnan, Q.~Sun, and J.~Zhou (2013).
\newblock Bivariate degradation analysis of products based on {Wiener}
  processes and copulas.
\newblock {\em Journal of Statistical Computation and Simulation\/}~{\em 83},
  1316--1329.

\bibitem[\protect\citeauthoryear{Park and Padgett}{Park and
  Padgett}{2005}]{ParkPadgett2005}
Park, C. and W.~J. Padgett (2005).
\newblock Accelerated degradation models for failure based on geometric
  \protect{Brownian} motion and gamma processes.
\newblock {\em Lifetime Data Analysis\/}~{\em 11}, 511--527.

\bibitem[\protect\citeauthoryear{Park and Staicu}{Park and
  Staicu}{2015}]{ParkStaicu2015}
Park, S.~Y. and A.-M. Staicu (2015).
\newblock Longitudinal functional data analysis.
\newblock {\em Stat\/}~{\em 4}, 212--226.

\bibitem[\protect\citeauthoryear{Peng}{Peng}{2016}]{Peng2016}
Peng, C.-Y. (2016).
\newblock Inverse {Gaussian} processes with random effects and explanatory
  variables for degradation data.
\newblock {\em Technometrics\/}~{\em 57}, 100--111.

\bibitem[\protect\citeauthoryear{Saxena and Goebel}{Saxena and
  Goebel}{2008}]{SaxenaGoebel2008}
Saxena, A. and K.~Goebel (2008).
\newblock {PHM08} challenge data set.
\newblock Technical report, NASA Ames Prognostics Data Repository, Moffett
  Field, CA.

\bibitem[\protect\citeauthoryear{Si, Yang, and Wu}{Si
  et~al.}{2017}]{SiYangWu2016}
Si, W., Q.~Yang, and X.~Wu (2017).
\newblock A distribution-based functional linear model for reliability analysis
  of advanced high strength dual-phase steels by utilizing material
  microstructure images.
\newblock {\em IIE Transactions on Quality and Reliability Engineering\/}~{\em
  49}, 863--873.

\bibitem[\protect\citeauthoryear{Si, Yang, Wu, and Chen}{Si
  et~al.}{2018}]{Sietal2018}
Si, W., Q.~Yang, X.~Wu, and Y.~Chen (2018).
\newblock Reliability analysis considering dynamic material local deformation.
\newblock {\em Journal of Quality Technology\/}~{\em 50}, xx--xx.

\bibitem[\protect\citeauthoryear{Steinberg}{Steinberg}{2016}]{DavidMSteinberg2016}
Steinberg, D.~M. (2016).
\newblock Industrial statistics: The challenges and the research.
\newblock {\em Quality Engineering\/}~{\em 28}, 45--59.

\bibitem[\protect\citeauthoryear{Sun, Liu, Li, and Liao}{Sun
  et~al.}{2016a}]{Sunetal2016a}
Sun, F., J.~Liu, X.~Li, and H.~Liao (2016a).
\newblock Reliability analysis with multiple dependent features from a
  vibration-based accelerated degradation test.
\newblock {\em Shock and Vibration, DOI:
  http://dx.doi.org/10.1155/2016/2315916\/}~{\em 16}, 1242.

\bibitem[\protect\citeauthoryear{Sun, Liu, Li, and Liao}{Sun
  et~al.}{2016b}]{Sunetal2016}
Sun, F., L.~Liu, X.~Li, and H.~Liao (2016b).
\newblock Stochastic modeling and analysis of multiple nonlinear accelerated
  degradation processes through information fusion.
\newblock {\em Sensors\/}~{\em 16}, 1242.

\bibitem[\protect\citeauthoryear{Tibshirani and Taylor}{Tibshirani and
  Taylor}{2011}]{TibshiraniTaylor2011}
Tibshirani, R.~J. and J.~Taylor (2011).
\newblock The solution path of the generalized lasso.
\newblock {\em Annals of Statistics\/}~{\em 39}, 1335--1371.

\bibitem[\protect\citeauthoryear{Wang and Zhu}{Wang and
  Zhu}{2008}]{WangZhu2008}
Wang, S. and J.~Zhu (2008).
\newblock Variable selection for model-based high-dimensional clustering and
  its application to microarray data.
\newblock {\em Biometrics\/}~{\em 64}, 440--448.

\bibitem[\protect\citeauthoryear{Wang and Xu}{Wang and Xu}{2010}]{WangXu2010}
Wang, X. and D.~Xu (2010).
\newblock An inverse \protect{Gaussian} process model for degradation data.
\newblock {\em Technometrics\/}~{\em 52}, 188--197.

\bibitem[\protect\citeauthoryear{Wang, Ye, Hong, and Tang}{Wang
  et~al.}{2018}]{Wangetal2017}
Wang, X., Z.-S. Ye, Y.~Hong, and L.-C. Tang (2018).
\newblock Analysis of field return data with failed-but-not-reported events.
\newblock {\em Technometrics, in press, DOI: 10.1080/00401706.2017.1292957\/}.

\bibitem[\protect\citeauthoryear{Wang and Zhu}{Wang and
  Zhu}{2017}]{WangZhu2017}
Wang, X. and H.~Zhu (2017).
\newblock Generalized scalar-on-image regression models via total variation.
\newblock {\em Journal of the American Statistical Association\/}~{\em 112},
  1156--1168.

\bibitem[\protect\citeauthoryear{Whitmore}{Whitmore}{1995}]{Whitmore1995}
Whitmore, G.~A. (1995).
\newblock Estimation degradation by a \protect{Wiener} diffusion process
  subject to measurement error.
\newblock {\em Lifetime Data Analysis\/}~{\em 1}, 307--319.

\bibitem[\protect\citeauthoryear{Whitmore, Crowder, and Lawless}{Whitmore
  et~al.}{1998}]{WhitmoreCrowderLawless1998}
Whitmore, G.~A., M.~J. Crowder, and J.~F. Lawless (1998).
\newblock Failure inference from a marker process based on a bivariate
  \protect{Wiener} model.
\newblock {\em Lifetime Data Analysis\/}~{\em 4}, 229--251.

\bibitem[\protect\citeauthoryear{Wu and Meeker}{Wu and
  Meeker}{2002}]{WuMeeker2002}
Wu, H. and W.~Q. Meeker (2002).
\newblock Early detection of reliability problems using information from
  warranty databases.
\newblock {\em Technometrics\/}~{\em 44}, 120--133.

\bibitem[\protect\citeauthoryear{Xie, King, Hong, and Yang}{Xie
  et~al.}{2018}]{Xieetal2016}
Xie, Y., C.~B. King, Y.~Hong, and Q.~Yang (2018).
\newblock Semi-parametric models for accelerated destructive degradation test
  data analysis.
\newblock {\em Technometrics, in press, DOI: 10.1080/00401706.2017.1321584\/}.

\bibitem[\protect\citeauthoryear{Xu, Hong, and Jin}{Xu
  et~al.}{2016}]{XuHongJin2015}
Xu, Z., Y.~Hong, and R.~Jin (2016).
\newblock Nonlinear general path models for degradation data with dynamic
  covariates.
\newblock {\em Applied Stochastic Models in Business and Industry\/}~{\em 32},
  153--167.

\bibitem[\protect\citeauthoryear{Xu, Hong, Meeker, Osborn, and Illouz}{Xu
  et~al.}{2017}]{Xuetal2017}
Xu, Z., Y.~Hong, W.~Q. Meeker, B.~E. Osborn, and K.~Illouz (2017).
\newblock A multi-level trend-renewal process for modeling systems with
  recurrence data.
\newblock {\em Technometrics\/}~{\em 59}, 225--236.

\bibitem[\protect\citeauthoryear{Yashchin}{Yashchin}{2012}]{Yashchin2012}
Yashchin, E. (2012).
\newblock Design and implementation of systems for monitoring lifetime data.
\newblock In H.-J. Lenz, W.~Schmid, and P.-T. Wilrich (Eds.), {\em Frontiers in
  Statistical Quality Control 10}, pp.\  171--195. Heidelberg: Physica-Verlag
  HD.

\bibitem[\protect\citeauthoryear{Ye and Chen}{Ye and Chen}{2014}]{YeChen2014}
Ye, Z.-S. and N.~Chen (2014).
\newblock The inverse {Gaussian} process as a degradation model.
\newblock {\em Technometrics\/}~{\em 56}, 302--311.

\bibitem[\protect\citeauthoryear{Yokoyama}{Yokoyama}{2016}]{MasahiroYokoyam2016}
Yokoyama, M. (2016).
\newblock A study on estimation of lifetime distribution with covariates under
  misspecification for baseline distribution.
\newblock {\em Engineering Letters\/}~{\em 24}, 195--201.

\bibitem[\protect\citeauthoryear{Yokoyama, Yamamoto, and Suzuki}{Yokoyama
  et~al.}{2015}]{Yokoyama2015}
Yokoyama, M., W.~Yamamoto, and K.~Suzuki (2015).
\newblock A study on estimation of lifetime distribution with covariates using
  online monitoring.
\newblock {\em Total Quality Science\/}~{\em 1}, 89--101.

\bibitem[\protect\citeauthoryear{Yuan and Lin}{Yuan and
  Lin}{2006}]{YuanLin2006}
Yuan, M. and Y.~Lin (2006).
\newblock Model selection and estimation in regression with grouped variables.
\newblock {\em Journal of the Royal Statistical Society, Series B\/}~{\em 68},
  49--67.

\bibitem[\protect\citeauthoryear{Zhang}{Zhang}{2015}]{Zhang2015}
Zhang, L. (2015).
\newblock {\em Big Data Analytics for {eMaintenance}: Modeling of
  high-dimensional data streams}.
\newblock Licentiate thesis, Lule{\aa} University of Technology, URL:
  https://www.diva-portal.org/smash/get/diva2:990005/FULLTEXT01.pdf.

\bibitem[\protect\citeauthoryear{Zhou, Li, and Zhu}{Zhou
  et~al.}{2013}]{ZhouLiZhu2013}
Zhou, H., L.~Li, and H.~Zhu (2013).
\newblock Tensor regression with applications in neuroimaging data analysis.
\newblock {\em Journal of the American Statistical Association\/}~{\em 108},
  540--552.

\bibitem[\protect\citeauthoryear{Zhou, Serban, and Gebraeel}{Zhou
  et~al.}{2011}]{ZhouSerbanGebraeel2011}
Zhou, R., N.~Serban, and N.~Gebraeel (2011).
\newblock Degradation modeling applied to residual lifetime prediction using
  functional data analysis.
\newblock {\em The Annals of Applied Statistics\/}~{\em 5}, 1586--1610.

\bibitem[\protect\citeauthoryear{Zhou, Serban, Gebraeel, and M{\"u}ller}{Zhou
  et~al.}{2014}]{Zhouetal2014}
Zhou, R., N.~Serban, N.~Gebraeel, and H.-G. M{\"u}ller (2014).
\newblock A functional time warping approach to modeling and monitoring
  truncated degradation signals.
\newblock {\em Technometrics\/}~{\em 56}, 67--77.

\bibitem[\protect\citeauthoryear{Zhu, Yashchin, and Hosking}{Zhu
  et~al.}{2014}]{ZhuYashchinHosking2014}
Zhu, Y., E.~Yashchin, and J.~Hosking (2014).
\newblock Parametric estimation for window censored recurrence data.
\newblock {\em Technometrics\/}~{\em 56}, 55--66.

\bibitem[\protect\citeauthoryear{Zin, Tin, and Hama}{Zin
  et~al.}{2016}]{ZinTinHama2016}
Zin, T.~T., P.~Tin, and H.~Hama (2016).
\newblock Reliability and availability measures for {Internet} of {Things}
  consumer world perspectives.
\newblock In {\em 2016 IEEE 5th Global Conference on Consumer Electronics},
  pp.\  1--2.

\bibitem[\protect\citeauthoryear{Zou and Hastie}{Zou and
  Hastie}{2005}]{ZouHastie2005}
Zou, H. and T.~Hastie (2005).
\newblock Regularization and variable selection via the elastic net.
\newblock {\em Journal of the Royal Statistical Society: Series B\/}~{\em 67},
  301--320.

\end{thebibliography}
\end{document}